\documentclass[amsmath,amssymb,amsbsy,reprint,pra,preprintnumbers,showpacs,superscriptaddress]{revtex4-2}
\usepackage{graphicx,color}
\usepackage{dcolumn}
\usepackage{bm}
\usepackage{braket}
\usepackage{mathtools}
\usepackage{ulem}
\usepackage[breaklinks,colorlinks=true,linkcolor=blue,urlcolor=blue,citecolor=blue]{hyperref}
\usepackage{times}

\newcommand{\txtb}[1]{\textcolor{blue}{#1}}

\begin{document}
\title{Orbital magnetic susceptibility of type-I, II, and III massless Dirac fermions in two dimensions}

\author{Tomonari Mizoguchi}
\affiliation{
Department of Physics,
University of Tsukuba, Tsukuba, Ibaraki 305-8571, Japan}
\email{mizoguchi@rhodia.ph.tsukuba.ac.jp}

\author{Hiroyasu Matsuura}
\affiliation{Department of Physics, University of Tokyo, Bunkyo-ku, Tokyo 113-0033, Japan}
\author{Masao Ogata}
\affiliation{Department of Physics, University of Tokyo, Bunkyo-ku, Tokyo 113-0033, Japan}
\affiliation{Trans-Scale Quantum Science Institute, University of Tokyo, Bunkyo-ku, Tokyo 113-0033, Japan}

\date{\today}

\begin{abstract}
We study the orbital magnetic susceptibility of 
tilted massless Dirac fermions in two dimensions.
It is well-known that 
the type-I massless Dirac fermions exhibit divergingly-large 
diamagnetic susceptibility, 
whereas less is known about the types II and III cases.
We first clarify that the orbital magnetic susceptibility 
is vanishing for the types II and III in the continuum model.
We then compare the 
three types of Dirac fermions 
for the lattice models.
We employ 
three tight-binding models with 
different numbers of Dirac points, 
all of which are two-band models defined on a square lattice.
For all three models, 
we find that the type-I Dirac fermions 
show the divergingly-large orbital diamagnetic 
susceptibility,
whereas the type-II Dirac fermions 
exhibit non-diverging paramagnetic susceptibility.
The type-III Dirac fermions exhibit diamagnetism but 
its susceptibility is small compared with the type-I case. 
\end{abstract}

\maketitle
\section{Introduction \label{sec:intro}}
The magnetic-field response of electrons in solids 
has been a fundamental and important issue. 
Under the periodic potentials of solids, 
the electronic states are described by the Bloch bands,
and their characteristic dispersion relations 
as well as the wavefunctions are known to
have a significant effect on the magnetic field response.
In non-magnetic materials with weak spin-orbit coupling, 
the magnetic susceptibility consists of two contributions, i.e., the spin and the orbital magnetic susceptibilities.
It has been revealed that 
the former is mainly governed by the density of state (DOS) around the Fermi energy,
whereas the latter has rich physical pictures, namely,
the interband effect plays an essential role~\cite{Blount1962,Fukuyama1969,Fukuyama1971,ogata2015} and 
some of such contributions can be attributed to the topology and geometry of the Bloch wavefunctions~\cite{Raoux2014,Raoux2015,Gao2015,Piechon2016,ogata2017,RHim2020,Ozaki2021,Ozaki2023,Ozaki2024}.

Dirac fermions are known to be a fertile ground of studying orbital magnetic susceptibility~\cite{Shoenberg1936,
Ganguli1941,McClure1956,McClure1960,Koshino2007,Koshino2007_2,Nakamura2007,Koshino2010,Koshino2011,Fuseya2015,Watanabe2021,Suetsugu2021,Fujiyama2022}. 
Among various systems, 
the type-I massless Dirac 
fermions in two dimensions,
which have linear dispersion and the 
ellipsoidal equi-energy surface around 
the band contact point, 
are known to exhibit characteristic orbital magnetic susceptibility~\cite{McClure1956,McClure1960,Koshino2007,Kobayashi2008,GomezSantos2011,Raoux2015,Ogata2016_2}.
To be more specific, 
when the chemical potential is right at the Dirac point,
the orbital magnetic susceptibility is diamagnetic and diverging.
Such a characteristic behavior has recently been observed experimentally in graphene~\cite{Vallejo2021}.
Meanwhile, little has been revealed for the 
overtitled and critically tilted Dirac fermions, which are dubbed 
the type-II (overtilted)~\cite{Soluyanov2015,Sun2015,Xu2015,Yu2016,Ruan2016} 
and type-III (critically tilted)~\cite{Volovik2016,Liu2018,Huang2018,
Fragkos2019,Milicevic2019,Kim2020,Chen2020,Jin2020,Mizoguchi2020_typeIII} Dirac fermions.
For these two types, 
the DOS is not vanishing 
even when the chemical potential is at the Dirac point,
which is in sharp contrast to 
the type-I Dirac fermions.

Several previous works have studied the orbital magnetic susceptibility 
of the tilted Dirac fermions 
by using the continuum model~\cite{Kobayashi2008,Schober2012,Suzumura2019}. 
It has been discussed that the orbital magnetic susceptibility positively diverges in the type-II Dirac fermions, i.e., the opposite way compared with the type-I~\footnote{See Supplemental Material of Ref.~\onlinecite{Schober2012}}.
In the present paper, we first study the continuum model where the Fermi surface forms straight lines in the type-II and type-III. 
We will show that the orbital magnetic susceptibility 
vanishes for these cases.

However, the continuum model has infinitely large Fermi surface and thus there will be some subtleties in the calculation for the real materials, such as the warping of the Fermi surfaces and/or bending at the boundary of the Brillouin zone.
One can avoid these subtleties by 
using lattice models, 
since the momentum-space summation is performed within the Brillouin zone,
and hence one can compare the results of all three types of the Dirac fermions 
in the common lattice models~\cite{Mizoguchi2022}. 
Therefore, in the present paper, 
we also study the orbital magnetic susceptibility
in the lattice models to compare the three types of Dirac fermions by varying the parameters.
We employ three tight-binding models with different numbers of Dirac points, 
all of which are two-band models 
defined on a square lattice.
For all three models, 
we find that the type-I Dirac fermions 
show the divergingly-large diamagnetic 
susceptibility,
whereas the type-II Dirac fermions 
exhibit non-diverging paramagnetic susceptibility.
The type-III Dirac fermions exhibit diamagnetism but its susceptibility is small compared with the type-I case. 

The rest of this paper is structured as follows. 
In Sec.~\ref{sec:model}, we explain our model and the formulation.
To be specific, we introduce the Dirac-type Hamiltonian with tilting in two dimensions
and show the theoretical formula for calculating the orbital magnetic susceptibility. 
In addition, before arguing the lattice model, we show the results for the continuum model for the tilted Dirac fermions in Sec.~\ref{sec:cont},
where we clarify that the type-II and type-III Dirac fermions exhibit
vanishing orbital magnetic susceptibility. 
In Sec.~\ref{sec:ham}, we introduce the square-lattice models.
We present three Hamiltonians hosting different number of the tilted Dirac cones. 
We also address the characteristic band structures of the three models, namely, the Dirac cones and the van Hove singularity. 
Our main results are 
presented in Sec.~\ref{sec:result}.
We numerically calculate the orbital magnetic susceptibility
and discuss its chemical potential dependence 
as well as the temperature dependence.
Section~\ref{sec:discussions} is devoted to the detailed discussion 
on the results in Sec.~\ref{sec:result}.
To be concrete, 
we discuss the role of the Dirac point for the type-II case. 
We also reveal that the paramagnetic peaks appearing away from the Dirac point can be explained 
by the intraband contribution. 
Finally, we present the summary of this paper in Sec.~\ref{sec:summary}.

\section{Orbital magnetic susceptibility and the 
continuum model with tilting~\label{sec:model}}

\subsection{Orbital magnetic susceptibility}
In general, orbital magnetic susceptibility is given by~\cite{Fukuyama1971}
\begin{align}
\chi = 
 \frac{e^2}{2\hbar^2}\frac{k_{\rm B}T}{V} \sum_{n, \bm{k}}  
\mathrm{Tr} \left[ v^x_{\bm{k}} {\mathcal G} v^y_{\bm{k}} {\mathcal G}
v^x_{\bm{k}} {\mathcal G} v^y_{\bm{k}} {\mathcal G} \right],
 \label{eq:fukuyama_formulaT}
\end{align}
where the summation is over the fermion Matsubara frequency, 
$\varepsilon_n = (2n+1)\pi k_{\rm B}T$ 
and 
the wave number $\bm k$.
Note that we neglect the spin degrees of freedom for simplicity. In other words, the results shown in this paper is the orbital magnetic susceptibility per spin degrees of freedom.
$V$ is the volume of the system 
(or the area for two dimensions). 
Tr is to take the trace over the Bloch bands and 
$\mathcal G$ is the abbreviation of the matrix form of thermal Green's function 
${\mathcal G}(\bm k, i\varepsilon_n)$ in the Bloch representation. 
The velocity $v^i_{\bm{k}}$ ($i  =x, y$) is defined in the matrix form as
$v^i_{\bm{k}} = \frac{\partial \mathcal{H}_{\bm{k}}}{\partial k_i}$,
$e$ is the elementary charge, 
$\hbar$ is the reduced Planck constant, 
and $k_{\rm B}$ is the Boltzmann constant.

In the following, we consider two-band models
whose Bloch Hamiltonians are generically given as
\begin{align}
\mathcal{H}_{\bm{k}} = h_{\bm{k}}^0 + \bm{h}_{\bm{k}} \cdot \bm{\sigma}, 
\label{eq:Ham_gen}
\end{align}
where $\bm{h}_{\bm{k}} = (h^x_{\bm{k}}, h^y_{\bm{k}},h^z_{\bm{k}})$, 
and $\bm{\sigma} = (\sigma_x, \sigma_y , \sigma_z)$ are the Pauli matrices. 
The dispersion relation of this Hamiltonian is given by
$\varepsilon_{\bm{k},\pm} = h^0_{\bm{k}} \pm |\bm{h}_{\bm{k}}|$.
Then the band touching occurs 
at which $|\bm{h}_{\bm{k}}| = 0$ is satisfied. $h_{\bm k}^0$ gives tilting.

In this Hamiltonian, we obtain thermal Green's function as
\begin{align}
    {\mathcal G}(\bm k, i\varepsilon_n) = \frac{i\tilde \varepsilon_n + 
    {\bm h}_{\bm k}\cdot \bm \sigma}{D},
\end{align}
with $i\tilde \varepsilon_n:= i \varepsilon_n - h_{\bm k}^0 +\mu$ 
($\mu$ is the chemical potential) and 
$D=(i\tilde \varepsilon_n)^2 - |\bm{h}_{\bm{k}}|^2$.
The velocities $v^i_{\bm{k}}$ ($i  =x, y$) are given by 
\begin{align}
    v^i_{\bm{k}} = \frac{\partial h_{\bm k}^0}{\partial k_i} + 
    \frac{\partial {\bm h}_{\bm k}}{\partial k_i} \cdot \bm \sigma .
\end{align}

\begin{figure*}[!tb]
\begin{center}
\includegraphics[clip,width = \linewidth]{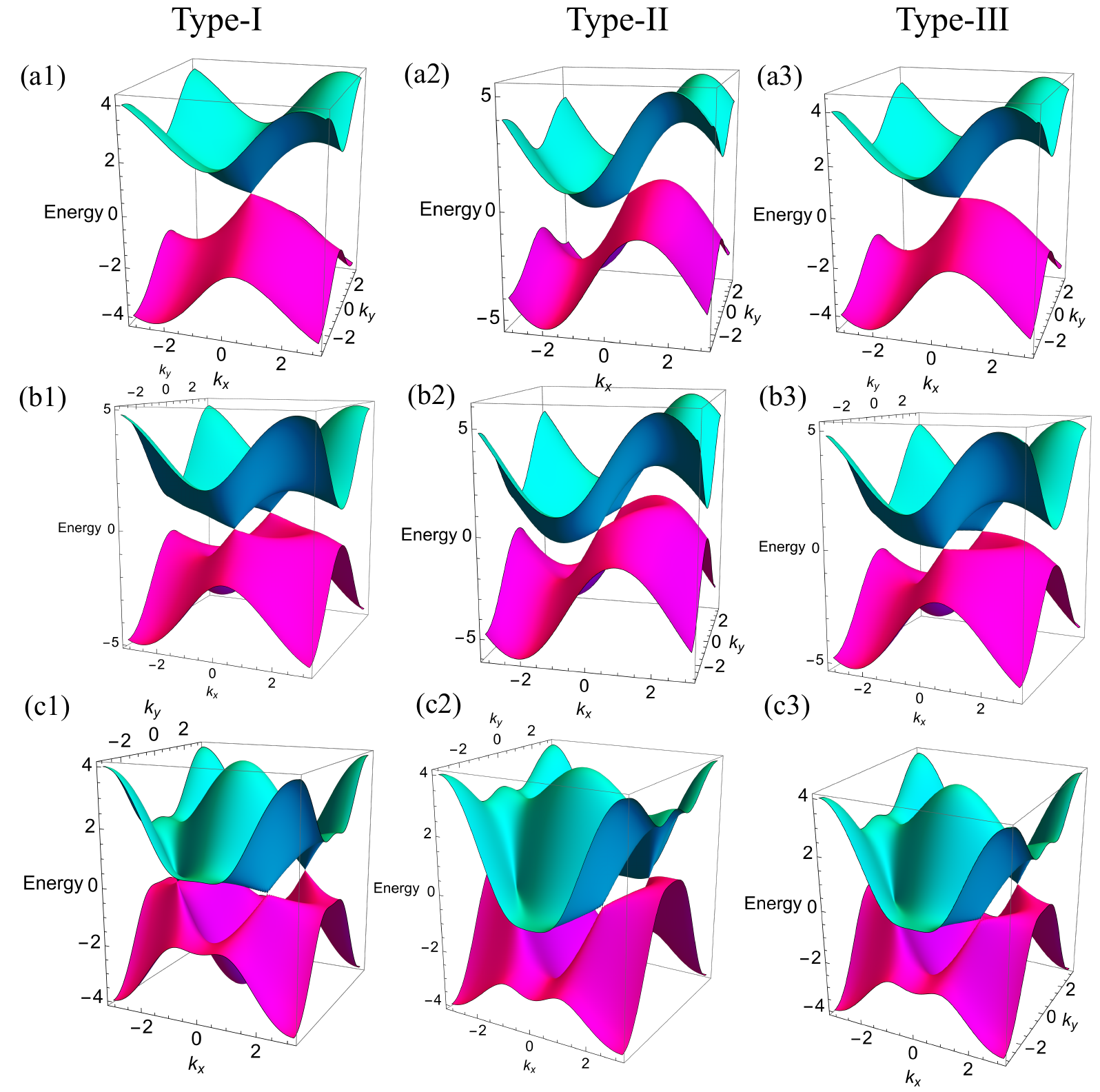}
\vspace{-10pt}
\caption{
The band structures of the models 
(i) for the panels labeled by (a), 
(ii) for the panels labeled by (b), and (iii) for the panels labeled by (c).
The panels labeled by 1, 2, and 3, are for the types I, II, and III Dirac fermions, respectively.
The parameters are described in Table~\ref{tab:table1}.
}
\label{fig:band}
\end{center}
\vspace{-10pt}
\end{figure*}
\subsection{Continuum model of massless Dirac Hamiltonian with tilting~\label{sec:cont}}
Before discussing the lattice model, we consider a single massless 
Dirac Hamiltonian in two-dimension with tilting. In this case, we use
\begin{align}
    \bm{h}_{\bm{k}} &= (v_{\rm F} \hbar k_x, v_{\rm F} \hbar k_y, 0), \cr
    h_{\bm k}^0 &= -\alpha v_{\rm F} \hbar k_x,
    \label{eq:Ham_2DDirac}
\end{align}
where $v_{\rm F}$ is the Fermi velocity and the parameter $\alpha$ represents 
the degree of tilting in the $k_x$-direction. 
The dispersion relation is given by 
$\varepsilon_{\bm{k},\pm} = -\alpha v_{\rm F}\hbar k_x\pm v_{\rm F} \hbar k$ 
with $k=|\bm k|$. 
The type I (II) corresponds to $0\le \alpha<1$ ($\alpha>1$) and 
The type III corresponds to $\alpha=1$ where the Dirac cone touches the $k_x$-$k_y$ plane.
In the case without tilting ($\alpha=0$), 
it is well known that at $T=0$ the orbital magnetic susceptibility 
$\chi$ has 
a delta function peak with negative sign at $\mu=0$. 
At finite temperature $\chi$ is proportional to $-1/T$~\cite{McClure1956,McClure1960}.

In the Hamiltonian (\ref{eq:Ham_2DDirac}), we obtain 
\begin{align}
    v^x_{\bm k} &= -\alpha v_{\rm F} \hbar + v_{\rm F} \hbar \sigma_x, \cr
    v^y_{\bm k} &= v_{\rm F} \hbar \sigma_y.
\label{eq:Velo_2DDirac}
\end{align}
Taking the trace in Eq.~(\ref{eq:fukuyama_formulaT}) and using the variables 
$x:= v_{\rm F}\hbar k_x$ and $y:= v_{\rm F}\hbar k_y$, we obtain 
the orbital magnetic susceptibility as
\begin{align}
    \chi =  e^2 v_{\rm F}^2 k_{\rm B}T \sum_n \iint \frac{dx dy}{4\pi^2}
    \left[ \frac{8y^2(x-\alpha i\tilde \varepsilon_n)^2}{D^4} - \frac{1-\alpha^2}{D^2} \right],
    \label{eq:Chi2DDirac1}
\end{align}
where $i\tilde \varepsilon_n:= i \varepsilon_n +\alpha x +\mu$ and 
$D=(i\tilde \varepsilon_n)^2 - x^2-y^2$.
Then, after we carry out the integration by parts, $\chi$ becomes
\begin{align}
    \chi = -\frac{e^2 v_{\rm F}^2}{6\pi^2} (1-\alpha^2) k_{\rm B}T 
    \sum_n \iint dx dy \frac{1}{D^2}.
    \label{eq:Chi2DDirac2}
\end{align}
From this, one may think that $\chi$ might change its sign when 
$\alpha$ becomes larger than 1 (i.e., the type II). 
However, it is not so simple because of the presence 
of the Fermi surface that extends to infinity in the $k_x$-$k_y$ plane 
in the type-II and type-III.

After a detailed calculation shown in Appendix~\ref{app}, we obtain
\begin{align}
    \chi =  \begin{cases} 
    \frac{e^2 v_{\rm F}^2}{6\pi} (1-\alpha^2)^{3/2} k_{\rm B}T \sum_n 
    \frac{1}{(i\varepsilon_n+\mu)^2}, & {\rm for }\ 0 \le \alpha<1,\cr 
    0, & {\rm for }\ 1 \le \alpha . \end{cases}
    \label{eq:Chi2DDirac3}
\end{align}
Rather surprisingly, this shows that $\chi$ is exactly zero in the type II Dirac Hamiltonian. 
[As we show in the following sections, 
$\chi$ has finite but small value in the type-II lattice Hamiltonian where the Fermi surface closes.] 
For $0 \le \alpha <1$, taking the summation over Matsubara frequency with use of 
$F(z) = 1/(e^{\beta z}+1)$ with $\beta = 1/(k_{\rm B}T)$, we obtain
\begin{align}
    \chi 
    &= -\frac{e^2 v_{\rm F}^2}{6\pi} (1-\alpha^2)^{3/2} \oint_C \frac{dz}{2\pi i} 
    F(z) \frac{1}{(z+\mu)^2} \cr
    &= \frac{e^2 v_{\rm F}^2}{6\pi} (1-\alpha^2)^{3/2} f'(0) \cr
    &= -\frac{e^2 v_{\rm F}^2 (1-\alpha^2)^{3/2}}{6\pi k_{\rm B}T} 
    \frac{e^{-\beta \mu}}{(e^{-\beta\mu}+1)^2}, 
    \label{eq:Chi2DDiracFinal}
\end{align}
where $f(\epsilon) = 1/(e^{\beta(\epsilon-\mu)}+1)$ is the Fermi-Dirac distribution function
and we have taken the residue at $z=\mu$. 
As $T$ goes to zero, $f'(0)$ approaches $-\delta(\mu)$ as in the well-known result.
On the other hand, when $\mu=0$ and at finite temperatures, 
$\chi$ is proportional to $-1/T$ as expected.

\begin{figure*}[!tb]
\begin{center}
\includegraphics[clip,width = \linewidth]{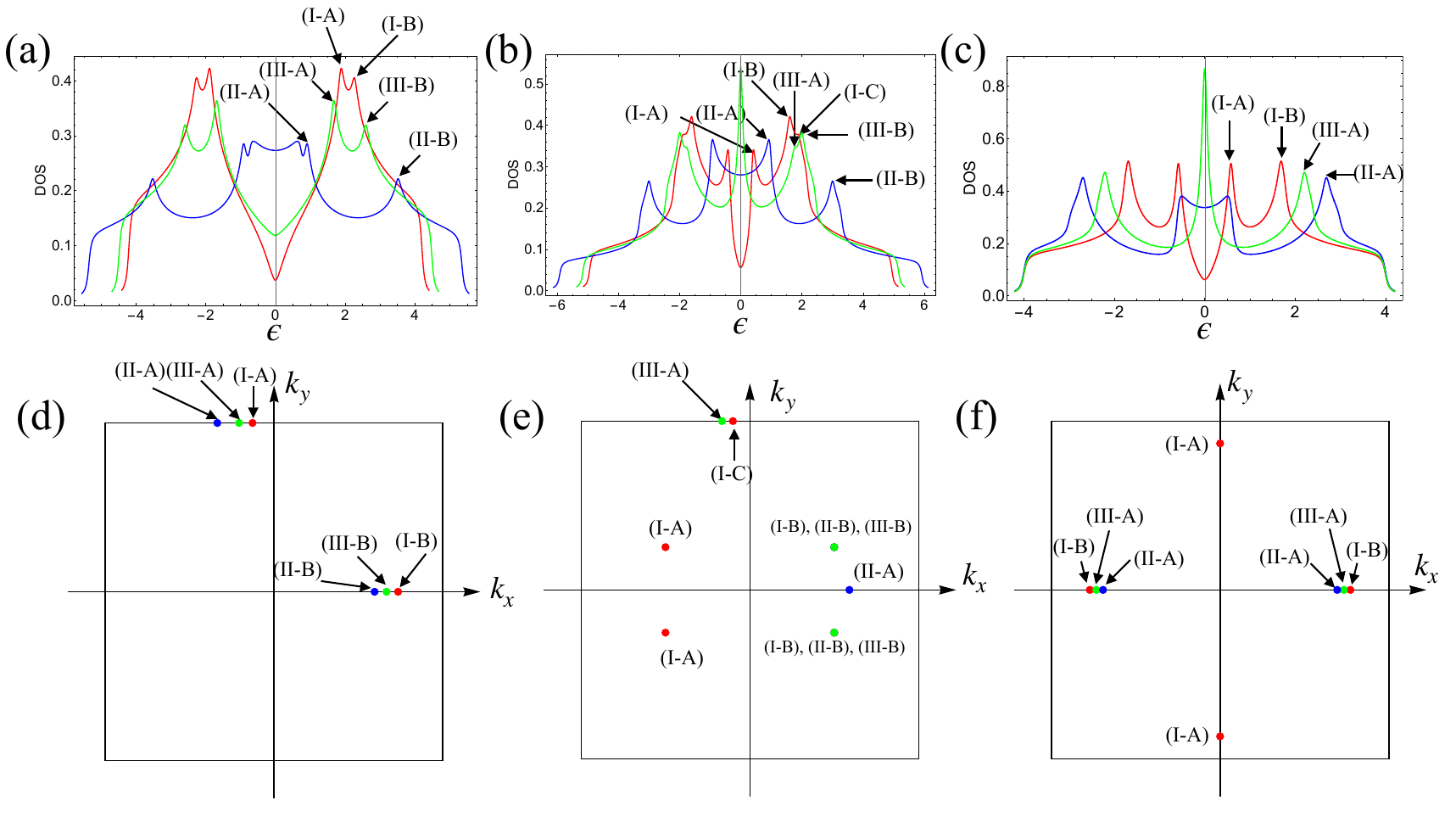}
\vspace{-10pt}
\caption{DOS for (a) the model (i), 
(b) the model (ii), and (c) the model (iii).
The positions of the van Hove singularities for 
(d) the model (i), 
(e) the model (ii), and (f) the model (iii).
For all panels, red, blue, and green lines and points denote the result for the types-I, II, and III, respectively.
}
\label{fig:dos}
\end{center}
\vspace{-10pt}
\end{figure*}

\begin{table*}[!tb]
\caption{
The tight-binding parameters for the models (i), (ii), and (iii), 
realizing the types-I, II, and III Dirac fermions. 
}
\begin{ruledtabular}
\begin{tabular}{cccc}
\textrm{}&
\textrm{$(t,t_x^0,m)$ for model (i)}&
\textrm{$(t,t_x^0,m)$ for model (ii)}&
\textrm{$(t,t^\prime,\lambda)$ for model (iii)}\\
\colrule
Type-I & (-1,-0.3,0.5) &(-1,-0.3,0.7) & (-1,-0.3,0.5)\\
Type-II & (-1,-1,0.5) & (-1,-1,0.7) & (-1,0.3,1.5)\\
Type-III & (-1,-0.5,0.5) & (-1,-0.5,0.7) & (-1,-0.3,1)\\
\end{tabular}
\end{ruledtabular}
\label{tab:table1}
\end{table*}
\section{Square-lattice tight-binding models for tilted Dirac Hamiltonian \label{sec:ham}}

Next, we consider the two-dimensional Dirac Hamiltonian on a square lattice.
The following three models are used:
\begin{itemize}
\item Model (i):
$h^0_{\bm{k}} = -2 t_x^0 \sin k_x$,
$h^x_{\bm{k}} =t \sin k_x$,
$h^y_{\bm{k}} = t \sin k_y$,
and
$h^z_{\bm{k}} = 2m(2-\cos k_x-\cos k_y)$.
\item Model (ii):
$h^0_{\bm{k}} = -2 t_x^0 \sin k_x$,
$h^x_{\bm{k}} =t \sin k_x$,
$h^y_{\bm{k}} =0$,
and
$h^z_{\bm{k}} = 2m(1-\cos k_x) + 2t \cos k_y$.
\item Model (iii):
$h^0_{\bm{k}} = 2 \lambda t^\prime (\cos k_x - \cos k_y)$,
$h^x_{\bm{k}} =2 t^\prime (\cos k_x - \cos k_y) $,
$h^y_{\bm{k}} = 0$,
and
$h^z_{\bm{k}} = 2t (\cos k_x + \cos k_y)$.
\end{itemize}
The models (i) and (ii) are the two-dimensional analogs of one employed in Ref.~\onlinecite{Udagawa2016},
and the model (iii) is based on Refs.~\onlinecite{Mizoguchi2020_typeIII,Mizoguchi2022}.
Note that, among various parameters in the models (i)-(iii), only $\lambda$ is dimensionless and all the others have the dimension of the energy.
Note also that we set the lattice constant to be unity.
The number of the Dirac points for the models (i), (ii), and (iii) is, respectively,
1, 2, and 4. 
To be more precise, 
the Dirac points of the models (i), (ii), and (iii) is, respectively,
$\bm{k} = (0,0)$, $\bm{k} = (0, \pm \pi/2)$, 
and $(\pm \pi/2, \pm \pi/2)$,
all of whose energy is 0.
 
We elaborate on 
the characteristics of 
the band structures of the models (i)-(iii).
Figure~\ref{fig:band} shows the band structures,
where the panels (a), (b), and (c) correspond to 
the models (i), (ii), and (iii), respectively,
and those labeled by
1, 2, and 3 correspond to 
the parameters realizing the Dirac fermions of the types I, II, and III, respectively.
The tight-binding parameters are listed in Table~\ref{tab:table1}.
As we have mentioned, 
we see Dirac cones 
whose numbers are 1, 2, and 4 for the models (i), (ii), and (iii), respectively.
By tuning a single parameter, 
i.e., $t_x^0$ for the models (i) and (ii) and $\lambda$ for the model (iii), 
one can control the 
tilting of the Dirac cones.
Among the three models, 
the model (iii) is characteristic 
in that, there is a direction where the dispersion is exactly flat for the type-III Dirac fermions, i.e., $k_x \pm k_y = \pm \pi$.
For the other two models, the dispersion is not exactly flat for the type-III Dirac fermions 
due to the higher-order terms of 
$\bm{k}$ around the Dirac points. 

Figures~\ref{fig:dos}(a), \ref{fig:dos}(b), and \ref{fig:dos}(c)
show the DOS for the models (i), (ii), and (iii), respectively, in the cases of types I-III. 
Here the DOS is defined as
\begin{align}
    \rho(\epsilon) = \frac{1}{V}
    \sum_{\bm{k},\eta = \pm}
    \frac{\Gamma}{\pi[(\epsilon-\varepsilon_{\bm{k},\eta})^2 + \Gamma^2]},
\end{align}
with 
$\Gamma$ being the damping rate. 
We set $\Gamma = 0.06$ for consistency 
to the calculations of the orbital magnetic susceptibility we show later.
For all three models, the DOS has a symmetry, $\rho(\epsilon) =\rho(-\epsilon)$, which reflects the symmetry of the band structures.
To be specific, $\varepsilon_{-\bm{k},-} = - \varepsilon_{\bm{k},+}$ for the models (i) and (ii), and
$\varepsilon_{\bm{k}+(\pi,\pi),-} = - \varepsilon_{\bm{k},+}$ for the model (iii). 
We also see that 
the type-I Dirac fermions 
have a vanishing DOS at the Dirac point (i.e., $\epsilon = 0$),
while the type-II Dirac fermions 
have a finite DOS.
For the type-III Dirac fermions, 
the DOS at the Dirac point 
exhibits a dip-like shape for the model (i), whereas
it exhibits a peak-like shape for the models (ii) and (iii).

Away from the Dirac point, 
we also see in Figs.~\ref{fig:dos}(a)-(c) that there exist several peaks
as denoted by the black arrows
(only those having the positive energies are marked).
They correspond to the van Hove singularities.
In Figs.~\ref{fig:dos}(d), \ref{fig:dos}(e), and
\ref{fig:dos}(f), we depict the corresponding momentum of each van Hove singularity, i.e., 
the saddle points of the dispersion, 
for the models (i), (ii), and (iii), respectively.

\section{Results~\label{sec:result}}
\begin{figure}[b]
\begin{center}
\includegraphics[clip,width = \linewidth]{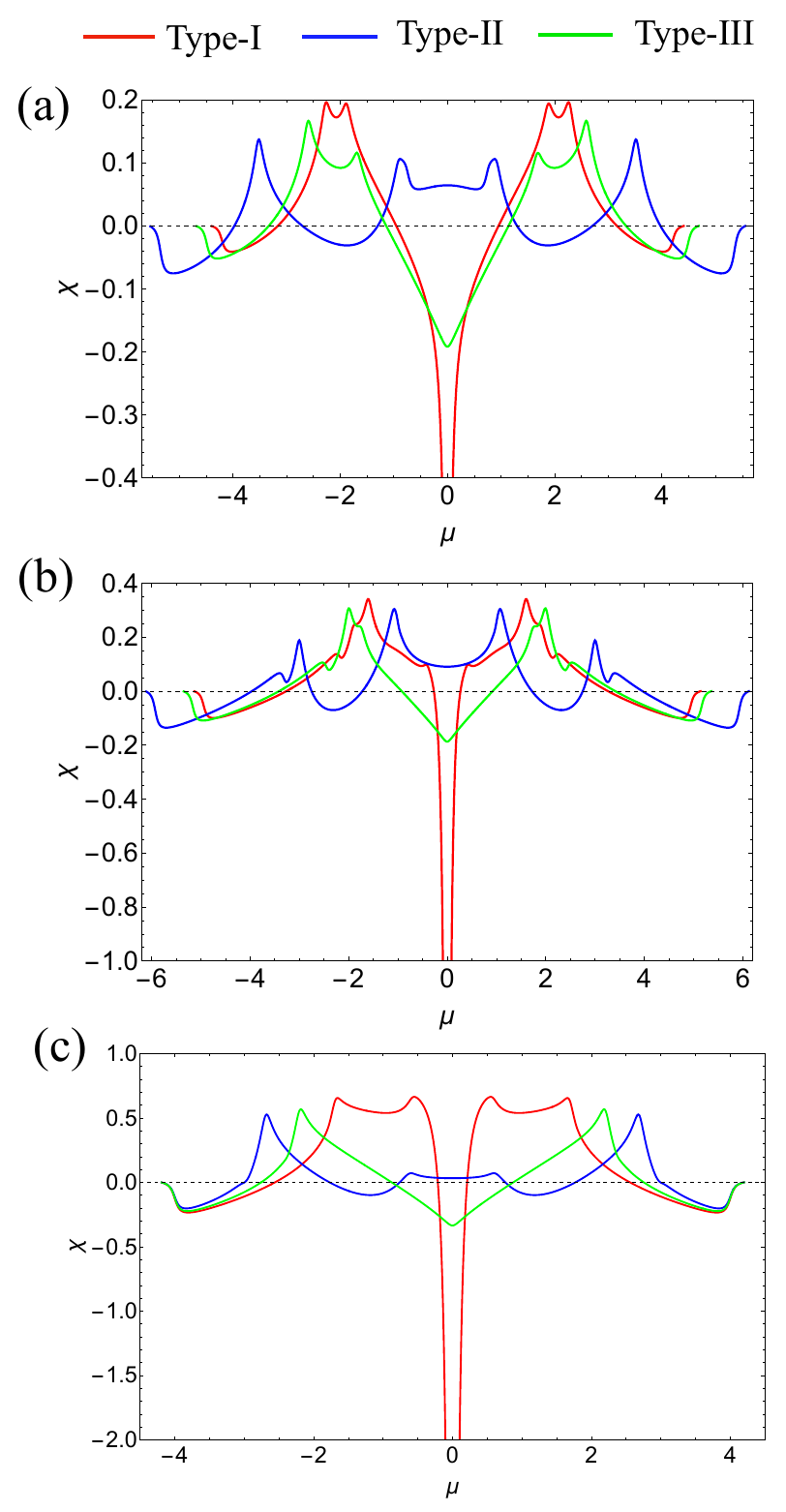}
\vspace{-10pt}
\caption{
Orbital magnetic susceptibility as a function of $\mu$ at $T=0$.
The panels (a), (b), and (c) correspond to 
the models (i), (ii), and (iii), respectively.
Note that the susceptibility is on the vertical axis in the unit of $\frac{e^2}{2 \pi \hbar^2}$.
}
\label{fig:orb}
\end{center}
\vspace{-10pt}
\end{figure}
In this section, we present the numerical results on the orbital magnetic susceptibility for the square-lattice models hosting Dirac cones. 

\subsection{Formulation for numerical calculation}
For the present models, we calculate 
the orbital magnetic susceptibility
numerically. 
As for the summation over 
the Matsubara frequency, we use the 
standard method to replace it to the energy integral with Fermi distribution function as
\begin{align}
\chi = 
-\frac{e^2}{2\pi \hbar^2}   
\int_{-\infty}^{\infty} d\epsilon f(\epsilon)  \mathrm{Im} [\Theta(\epsilon)],
\end{align}
with 
\begin{align} 
 \Theta(\epsilon)= 
 \frac{1}{V}\sum_{\bm{k}}  
\mathrm{Tr} 
[v^x_{\bm{k}} G_{\bm{k}}(\epsilon)v^y_{\bm{k}} G_{\bm{k}}(\epsilon)v^x_{\bm{k}} G_{\bm{k}}  (\epsilon)v^y_{\bm{k}} G_{\bm{k}}(\epsilon)].
 \label{eq:fukuyama_formula}
\end{align}
Here $G_{\bm{k}}(\epsilon) = 
(\epsilon+i \Gamma - \mathcal{H}_{\bm{k}})^{-1}$
is the retarded Green’s function.
Note that $\Gamma$ corresponds to the damping rate.
Note also that $\frac{\partial^2 \mathcal{H}_{\bm{k}}}{\partial k_x \partial k_y} = 0$ holds for the models (i)-(iii).
Several groups
have shown that the orbital magnetic susceptibility 
in the lattice model has the form~\cite{GomezSantos2011}
\begin{align} 
 \tilde\Theta(\epsilon)= 
 \frac{1}{V}&\sum_{\bm{k}}  
\mathrm{Tr} 
\biggl[v^x_{\bm{k}} G_{\bm{k}}(\epsilon)v^y_{\bm{k}} G_{\bm{k}}(\epsilon)v^x_{\bm{k}} G_{\bm{k}}  (\epsilon)v^y_{\bm{k}} G_{\bm{k}}(\epsilon) \cr
&+\frac{1}{2} \bigl\{ G_{\bm{k}}(\epsilon) v^{x}_{\bm{k}} G_{\bm{k}} v^{y}_{\bm{k}}
+G_{\bm{k}}(\epsilon) v^{y}_{\bm{k}} G_{\bm{k}} v^{x}_{\bm{k}} \bigr\} 
G_{\bm{k}} v^{xy}_{\bm{k}}\biggr], \cr
 \label{eq:Gomez_formula}
\end{align}
or~\cite{Raoux2015}
\begin{align} 
 \tilde\Theta(\epsilon)= 
 \frac{1}{V}\sum_{\bm{k}}  
&\mathrm{Tr} 
\biggl[\frac{2}{3} v^x_{\bm{k}} G_{\bm{k}}(\epsilon)v^y_{\bm{k}} G_{\bm{k}}(\epsilon)v^x_{\bm{k}} G_{\bm{k}}  (\epsilon)v^y_{\bm{k}} G_{\bm{k}}(\epsilon) \cr
&- \frac{2}{3} v^x_{\bm{k}} G_{\bm{k}}(\epsilon)v^x_{\bm{k}} G_{\bm{k}}(\epsilon)v^y_{\bm{k}} G_{\bm{k}}  (\epsilon)v^y_{\bm{k}} G_{\bm{k}}(\epsilon) \cr
&+\frac{1}{6} \bigl\{ v^{xx}_{\bm{k}} G_{\bm{k}}(\epsilon)v^{yy}_{\bm{k}} G_{\bm{k}}
-v^{xy}_{\bm{k}} G_{\bm{k}}(\epsilon)v^{xy}_{\bm{k}} G_{\bm{k}} \bigr\} \biggr], 
 \label{eq:Piechon_formula}
\end{align}
instead of $\Theta(\epsilon)$ in Eq.~(\ref{eq:fukuyama_formula}), where 
$v^{ij}_{\bm{k}}=\frac{\partial^2 \mathcal{H}_{\bm{k}}}{\partial k_i \partial k_j}$. 
In the present case where
$\frac{\partial^2 \mathcal{H}_{\bm{k}}}{\partial k_x \partial k_y} = 0$ holds, 
it is apparent that the results using Eq.~(\ref{eq:Gomez_formula}) is the same as those using 
Eq.~(\ref{eq:fukuyama_formula}). 
We also find that Eq.~(\ref{eq:Piechon_formula})  
is equivalent to Eq.~(\ref{eq:fukuyama_formula}) 
in the present case as far as the boundary contributions in 
the integration by parts vanish.
(See the appendix B for details.) 
Therefore, the above three formulae for 
$\tilde\Theta(\epsilon)$ give the same results.

\subsection{Numerical results}
Figures~\ref{fig:orb}(a), \ref{fig:orb}(b), and \ref{fig:orb}(c) show the $\mu$ dependence of the orbital magnetic susceptibility for the models (i), (ii), and (iii), respectively. 
Here we set $T=0$ and $\Gamma = 0.06$. 
Focusing on the Dirac point ($\mu = 0$), all three models exhibit
qualitatively similar behavior.
Specifically, we see that the type-I Dirac fermions exhibit divergingly large diamagnetic susceptibility, which coincides with 
the result in the previous section for the continuum model and
the previous works~\cite{McClure1956,McClure1960,Koshino2007,Kobayashi2008,GomezSantos2011,Raoux2015,Ogata2016_2}.
Meanwhile, the type-II Dirac fermions 
do not show divergingly large susceptibility.
It has a small positive value and 
the 
$\mu$ dependence around the Dirac point is very small, which is in sharp contrast to the 
type-I case where the steep drop of $\chi$ around the Dirac point is seen. 
Note that in the continuum model in Sec.~\ref{sec:cont}, the orbital magnetic susceptibility vanishes, meaning that the finite values of $\chi$ 
obtained here originate from the lattice model.
As for the type-III case, we see the diamagnetic susceptibility, but its value is 
much smaller than that for the type-I.
Comparing among three models, we see that the ratio between the orbital magnetic susceptibility 
of the type-III and the type-I becomes smaller as the number of the Dirac cones increases. 

Away from the Dirac points, 
we see several peaks of $\chi$,
all of which are positive-signed (i.e., paramagnetic).
We elaborate on the origin of these peaks in the next section.

\begin{figure}[t]
\begin{center}
\includegraphics[clip,width = \linewidth]{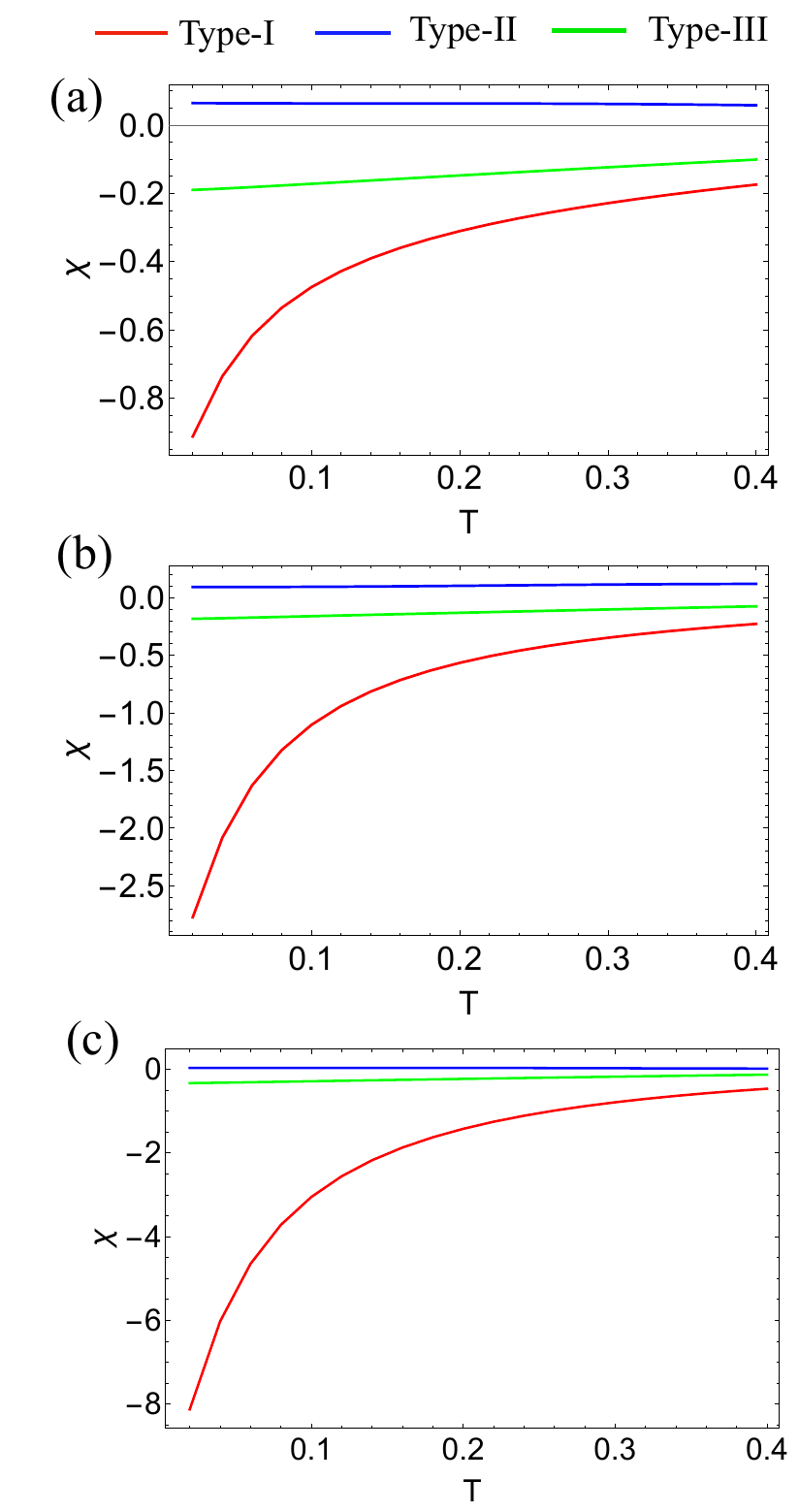}
\vspace{-10pt}
\caption{
Orbital magnetic susceptibility as a function of $T$ at $\mu = 0$.
The panels (a), (b), and (c) correspond to 
the models (i), (ii), and (iii), respectively.
$\chi$ in the vertical axis is in the unit of 
$\frac{e^2}{2 \pi \hbar^2}$.
}
\label{fig:orb_temp}
\end{center}
\vspace{-10pt}
\end{figure}

Figure~\ref{fig:orb_temp} 
shows the temperature 
dependence of $\chi$ with $\mu = 0$.
Again, qualitative behavior 
is common among the three models. 
We see that only the type-I Dirac fermions exhibit
steep temperature dependence, namely,
the diamagnetic susceptibility becomes larger as $T \rightarrow 0$.
In contrast, $\chi$ exhibits a small temperature dependence 
for the types-II and III Dirac fermions.

Summarizing these results,
we find that only the type-I Dirac fermions show characteristic divergingly large diamagnetism at the Dirac points,
which has a sharp dependence on $\mu$ and $T$,
whereas the type-II Dirac fermions show 
small paramagnetism which is rather insensitive to $\mu$ and $T$.
The type-III fermions show the diamagnetism which is moderately dependent on $\mu$ but rather insensitive to $T$.

\section{Discussions \label{sec:discussions}}

\subsection{Role of Dirac points in type-II case}

\begin{figure}[tb]
\begin{center}
\includegraphics[clip,width = 0.8\linewidth]{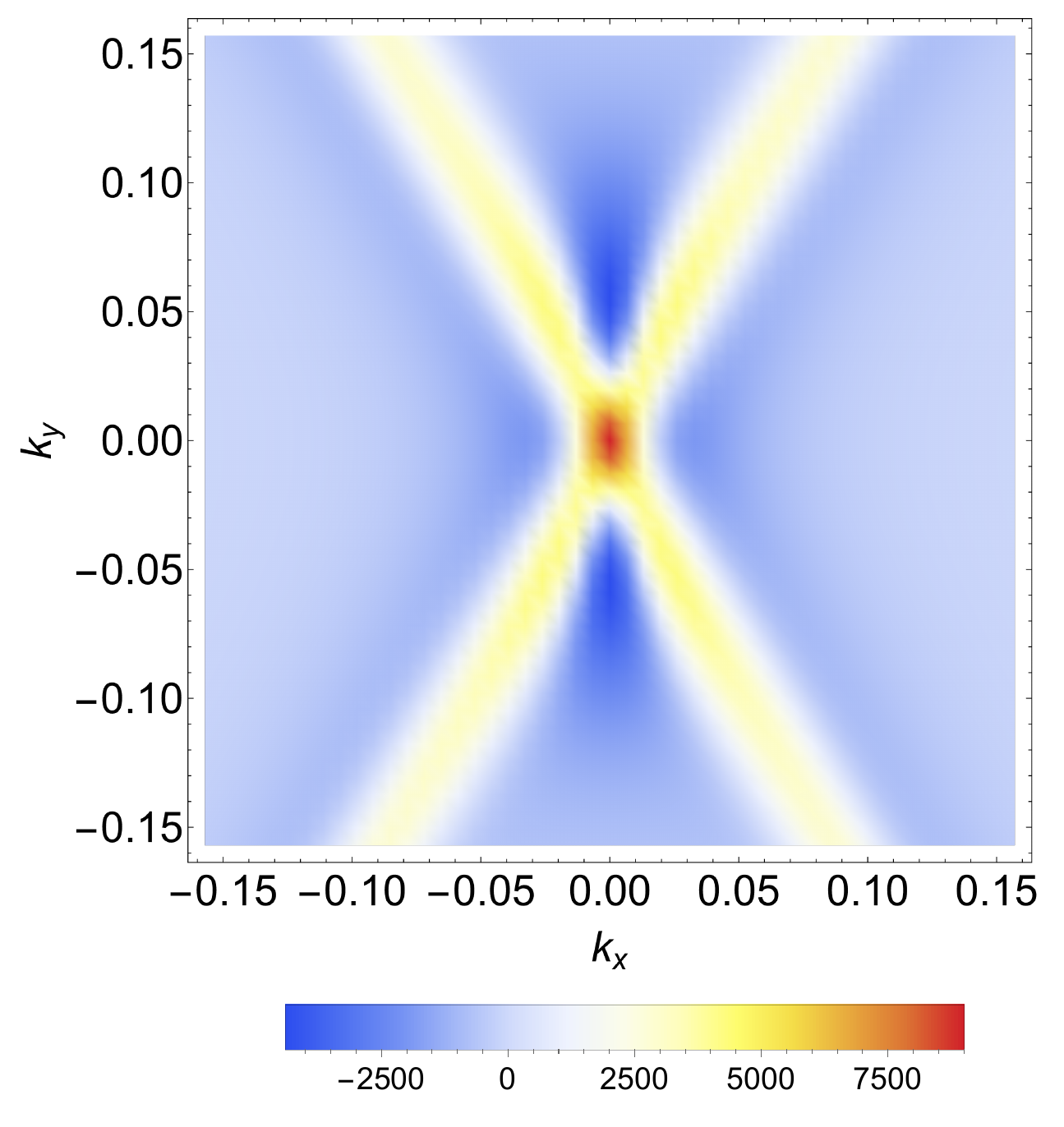}
\vspace{-10pt}
\caption{The map of 
$\bar{\chi}_{\bm{k}}$ for the continuum model of Eq.~(\ref{eq:Ham_2DDirac}).
We set $\alpha = 2$, $v_{\rm F} = 1$, $\mu =0$, and $\Gamma = 0.06$.
}
\label{fig:mom_resolved_cont}
\end{center}
\vspace{-10pt}
\end{figure}

\begin{figure*}[t]
\begin{center}
\includegraphics[clip,width = 0.95\linewidth]{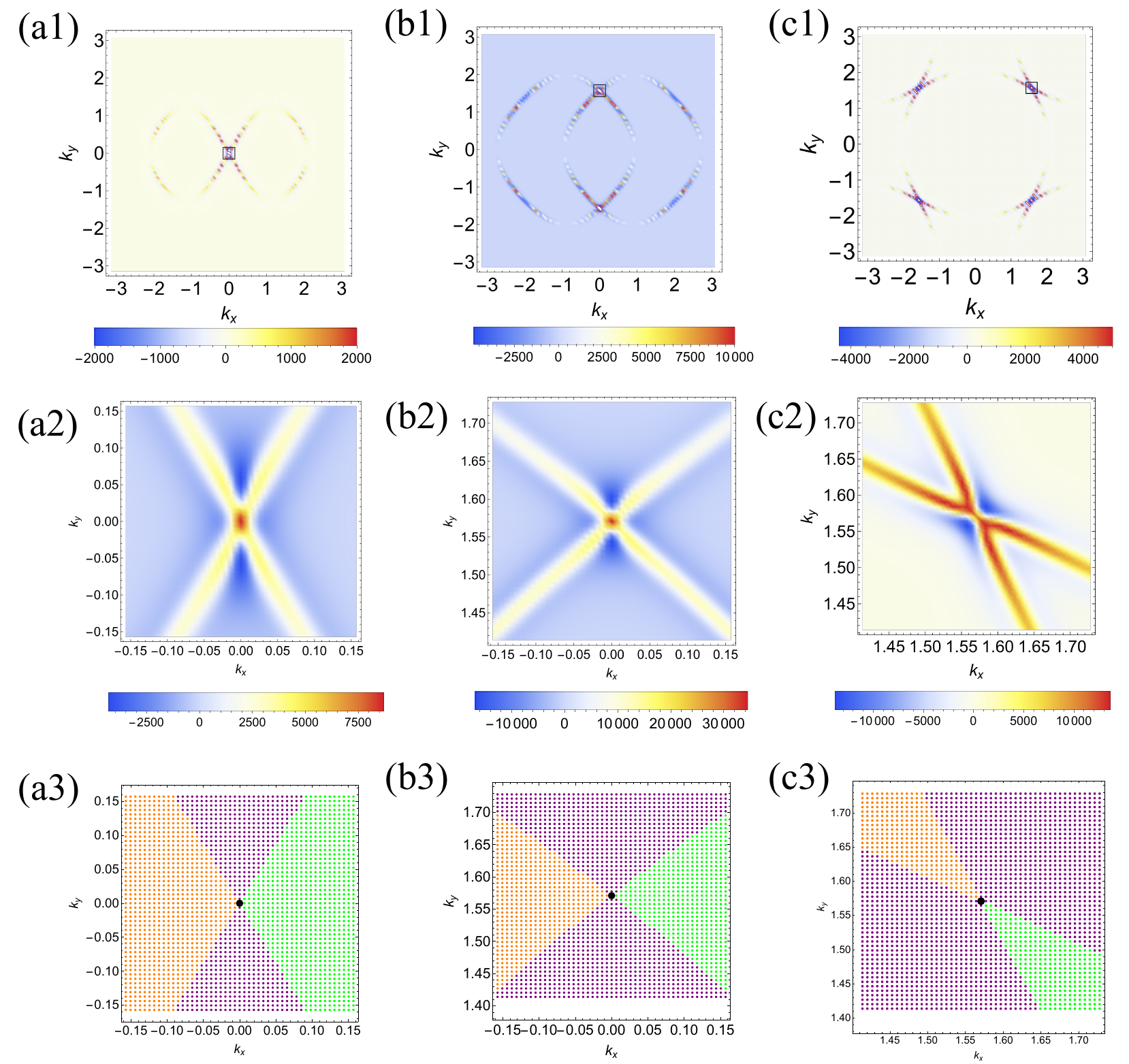}
\vspace{-10pt}
\caption{The map of 
$\bar{\chi}_{\bm{k}}$ of Eq.~(\ref{eq:chibar}) for the type-II case 
at $\mu = 0$.
The panels (a), (b), and (c) are for the models (i), (ii), and (iii), respectively.
The upper panels are the map for the entire Brillouin zone.
The middle panels are the zoom-up near a Dirac point,
i.e., $\bm{k} = (0,0)$ for (a), $(0,\pi/2)$ for (b), and  $(\pi/2,\pi/2)$ for (c). 
The lower panels represent the classification of the regions near the Dirac point with respect to the occupation of the bands. The orange dots represent the region where both upper and the lower bands are occupied.
The green dots represent the region where both upper and the lower bands are unoccupied.
The purple dots represent the region where only the lower band is occupied. 
}
\label{fig:mom_resolved}
\end{center}
\vspace{-10pt}
\end{figure*}

Unlike the type-I case, in the type-II case,
the Fermi surface extends the region away from the Dirac points. 
Thus, here we argue whether 
the Dirac points play essential roles  
for the orbital magnetic susceptibility 
when the chemical potential is at the Dirac point. 
To this end, we compute the momentum-resolved contributions to the orbital magnetic susceptibility~\footnote{Note that $\bar{\chi}_{\bm{k}}$ is not divided by the volume.}:
\begin{align}
\bar{\chi}_{\bm{k}} 
= - \int_{-\infty}^\infty d\epsilon f(\epsilon) \mathrm{Im}[\theta_{\bm{k}}(\epsilon)], \label{eq:chibar}
\end{align}
with 
\begin{align}
\theta_{\bm{k}}(\epsilon) = \mathrm{Tr} 
[v^x_{\bm{k}} G_{\bm{k}}(\epsilon)v^y_{\bm{k}} G_{\bm{k}}(\epsilon)v^x_{\bm{k}} G_{\bm{k}}  (\epsilon)v^y_{\bm{k}} G_{\bm{k}}(\epsilon)].
\end{align}

First, we examine how the orbital magnetic susceptibility vanishes in the continuum model in the type-II case.
We calculate $\bar{\chi}_{\bm k}$ using Eqs.~(\ref{eq:Ham_2DDirac}) and (\ref{eq:Velo_2DDirac}) with $\alpha=2$ at $\mu = 0$ with $\Gamma=0.06$.
As shown in Fig.~\ref{fig:mom_resolved_cont},
we can see that there are delta-function-like positive peaks maximized at the Dirac point along $\alpha\cos\theta\pm 1=0$
[$\theta = \tan^{-1}(k_y/k_x)$], which correspond to the Fermi surfaces of the continuum model. 
On the other hand, 
in the region where both $\alpha\cos\theta+1>0$ and $\alpha\cos\theta-1<0$ hold
(i.e., the region where only the lower band is occupied), there are negative contributions. 
As pointed out in Eq.~(\ref{eq:Chi2DDirac3}), 
the orbital magnetic susceptibility exactly 
vanishes after performing the integration over $\bm{k}$.
Therefore, we expect that the cancellation occurs between the large positive contribution at the Fermi surface and the negative contributions from the other region in the continuum model.

Let us proceed to the lattice models. 
In Fig.~\ref{fig:mom_resolved}, we plot $\bar{\chi}_{\bm{k}}$
for the type-II case at $\mu = 0$.
The upper panels are the map for the entire Brillouin zone.
We see that the main contributions originate from the Fermi surface. 
The middle panels are the zoom-up near a Dirac point. 
We see that $\bar{\chi}_{\bm{k}}$ sharply oscillates
along the direction perpendicular to the tilting of the cone. 
To be specific, $\bar{\chi}_{\bm{k}}$ has a large positive value right at the Dirac point, 
whereas it has a large positive value along the Fermi surfaces and
a large negative value in the two areas out of four areas away from the Dirac point.
For closer look
at this, in the lower panels, 
we classify the momentum space
with respect to the band occupation.
There are classified into three types as indicated by the colors.
Specifically, the orange (green) dots represent the region where both upper and the lower bands are occupied (unoccupied),
and the purple dots represent the region where only the lower band is occupied. 
As mentioned above, we see that the positive contribution comes 
from the Fermi surface, 
maximized at the Dirac point, 
while the large negative value comes from the region represented by the purple dots.
These behaviors are qualitatively the same behavior as in the continuum model shown in Fig.~\ref{fig:mom_resolved_cont}.
However, the exact cancellation does not occur in the lattice models.
As a result of this subtle cancellation, 
the susceptibility for the type-II case at $\mu = 0$ is moderately paramagnetic [Fig.~\ref{fig:orb}].
This is in contrast with the type-I Dirac fermions that contribute divergingly to the diamagnetic susceptibility.

\subsection{Comparison to the Landau-Peierls formula}
In Sec.~\ref{sec:intro}, we have addressed the importance of the interband effect in the orbital magnetic susceptibility.
In this regard, it is worth interpreting our results in terms of the intraband versus interband contributions. 
To extract the intraband contribution in the present results,
we employ the Landau-Peierls formula~\cite{Landau1930,Peierls1933} to calculate the orbital magnetic susceptibility that purely comes from the intraband contribution:
\begin{align}
&\chi^{\mathrm{LP}} = \frac{e^2}{12 \hbar^2 V} \notag \\
&\times  
\sum_{\bm{k},\eta = \pm}
\left\{ \left(\frac{\partial^2 \varepsilon_{\bm{k},\eta}}{\partial k_x^2}\right)\left(\frac{\partial^2 \varepsilon_{\bm{k},\eta}}{\partial k_y^2}\right) -\left( \frac{\partial^2 \varepsilon_{\bm{k},\eta}}{\partial k_x  \partial k_y} \right)^2 \right\}\frac{\partial f(\varepsilon_{\bm{k},\eta})}{\partial \epsilon}.
\end{align}
We note that 
the damping rate $\Gamma$ is 
neglected in this formula, 
hence we cannot 
compare the results for $\chi^{\mathrm{LP}}$ 
with $\chi$ presented in Fig.~\ref{fig:orb} in a quantitative manner.
Nevertheless, we can extract the features of the $\mu$ dependence of $\chi$, 
as we will show later. 

\begin{figure*}[tb]
\begin{center}
\includegraphics[clip,width = \linewidth]{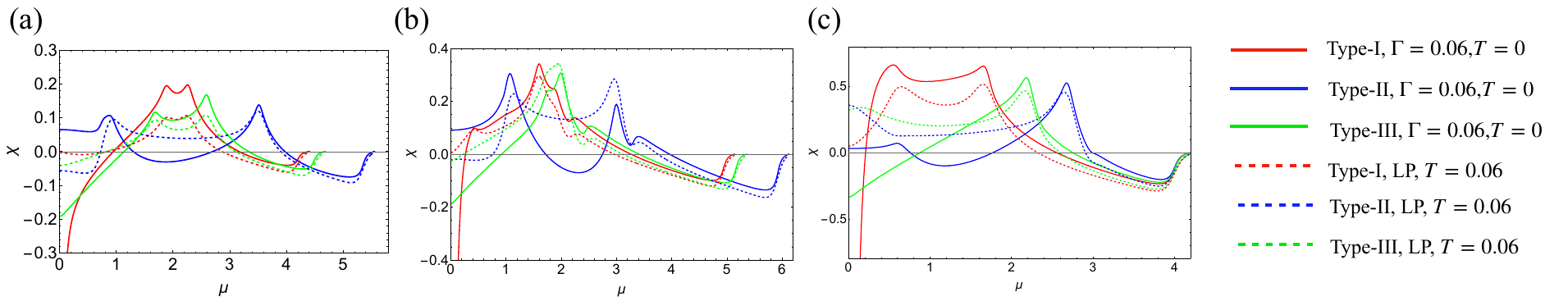}
\vspace{-10pt}
\caption{
Comparison between 
the total orbital magnetic susceptibility
at $T=0$, $\Gamma = 0.06$ (represented by the solid lines) and the 
Landau-Peierls susceptibility at $T = 0.06$ (represented by the dashed lines).
Panels (a), (b), and (c) are for the models (i), (ii), and (iii), respectively.
Note that the susceptibility is on the vertical axis in the unit of $\frac{e^2}{2 \pi \hbar^2}$.
}
\label{fig:LP}
\end{center}
\vspace{-10pt}
\end{figure*}

Figure~\ref{fig:LP} shows the $\chi$ and $\chi^{\rm LP}$
as functions of $\mu$ with $\mu \geq 0$.
Note that we set $T= 0.06$ in calculating $\chi^{\rm LP}$, so that we can compare $\chi^{\rm LP}$ with $\chi$ at the finite $\Gamma$ in an ad hoc manner.
We see that the paramagnetic peaks of $\chi$ can be accounted for by $\chi^{\rm LP}$. 
As mentioned in \txtb{Sec.~\ref{sec:ham}}, the peak positions correspond to the van Hove singularities.
Note that the emergence 
of the paramagnetic peaks at the van Hove singularities coincides with the various two-dimensional tight-binding models such as the single-orbital square lattice~\cite{Raoux2015,Ogata2016} and the honeycomb lattice~\cite{Raoux2015,GomezSantos2011,Ogata2016_2}.

We also see that $\chi^{\rm LP}$ (dashed lines
in Fig.~\ref{fig:LP}) are finite for  $\mu \sim 0$,
which does not reproduce the behavior of $\chi$
near $\mu = 0$
for all three types of the Dirac cones,
meaning that the interband contribution plays an 
essential role, not for the type-I but also for types-II and III. 

\section{Summary \label{sec:summary}}
We have investigated 
the orbital magnetic susceptibility for the two-dimensional massless Dirac fermions in the continuum model and the lattice models,
paying attention to the comparison among types I, II, and III. 
For studying the lattice models, 
we have employed three different models, 
all of which are 
defined on a square lattice and described by the 
$2 \times 2$ Hamiltonian in the momentum space. 

For the continuum model,
as is well-known, $\chi$ is diamagnetic and diverging for the type-I, but is proportional to $(1-\alpha)^{3/2}$ (i.e., the divergence is weaken as the tilting increases). 
In contrast to this, 
we have clarified that $\chi$ vanishes for the types II and III. 

We then compare this behavior with the lattice models. 
We have found that, in the case where the chemical potential is at the Dirac point, 
 the type-I Dirac fermions exhibit divergingly large diamagnetic susceptibility, the type-II Dirac fermions exhibit finite but small paramagnetic susceptibility, 
 and the type-III Dirac fermions exhibit diamagnetic susceptibility but 
 the divergence is much weaker. 
This behavior is different from the continuum model, which is attributed to the lattice.
As for the temperature dependence, only type-I Dirac fermions show strong temperature dependence, whereas 
types-II and III are almost temperature-independent.
The fact that the qualitative behaviors are in common among the three models indicates that the results are universal among the Dirac fermions for the lattice models. 
For the type-II case, we have investigated the momentum-resolved contributions 
to the orbital magnetic susceptibility,
and have found that the contribution near 
the Dirac points sharply oscillates, 
resulting in a small contribution.
Thus, the moderate paramagnetic susceptibility originates from the Fermi surface away from the Dirac points. 

We have also found that, away from $\mu = 0$ where Dirac points exist, 
there are several positive-sign peaks of the orbital magnetic susceptibility as a function of $\mu$.
They correspond to the van Hove singularity, 
and can be accounted for 
the Landau-Peierls-type 
intraband contributions.

\acknowledgements
T.M. thanks Nobuyuki Okuma 
for fruitful discussions.
This work is supported by 
JSPS KAKENHI, Grant 
No.~JP23K03243 and JP23K03274.

\appendix
\section{Orbital magnetic susceptibility in the massless Dirac Hamiltonian with tilting \label{app}}

In this appendix, we show some details of the calculations of $\chi$ 
in the two-dimensional Dirac Hamiltonian with tilting. 
Since $D=(i\tilde \varepsilon_n)^2 - x^2-y^2$ with $ i \tilde{\varepsilon}_n = i\varepsilon_n +\alpha x + \mu $, 
the following relation holds:
\begin{align}
    \frac{\partial}{\partial x} \frac{1}{D^3} = \frac{6}{D^4} (x-\alpha i\tilde \varepsilon_n).
    \label{eq:Trick1}
\end{align}
We can use this relation in the first term of Eq.~(\ref{eq:Chi2DDirac1}) to
perform the integration by parts. Then, we obtain
\begin{align}
    \chi =  e^2 v_{\rm F}^2 k_{\rm B}T \sum_n \iint \frac{dx dy}{4\pi^2}
    \left[ -\frac{4y^2(1-\alpha^2)}{3D^3} - \frac{1-\alpha^2}{D^2} \right].
    \label{eq:Chi2DDiracApp1}
\end{align}
Note that we have used 
\begin{align}
    \frac{4y^2(x- \alpha i\tilde\varepsilon_n)}{3D^3}\bigg|_{x=\pm \infty} = 0, 
\end{align}
which holds even when $\alpha= 1$. 
Furthermore, using the relation 
\begin{align}
    \frac{\partial}{\partial y} \frac{1}{D^2} = \frac{4y}{D^3},
\end{align}
Eq.~(\ref{eq:Chi2DDiracApp1}) becomes
\begin{align}
    \chi =  e^2 v_{\rm F}^2 k_{\rm B}T \sum_n \iint \frac{dx dy}{4\pi^2}
    \left[ \frac{1-\alpha^2}{3D^2} - \frac{1-\alpha^2}{D^2} \right],
    \label{eq:Chi2DDiracApp2}
\end{align}
which leads to Eq.~(\ref{eq:Chi2DDirac2}) in the main text. 

Next, we perform the $x$ and $y$ integrals. Using the polar coordinate in two-dimension,
we put $x=p\cos\theta$ and $y=p\sin\theta$. Then, Eq.~(\ref{eq:Chi2DDirac2}) becomes
\begin{align}
    \chi = -\frac{e^2 v_{\rm F}^2}{6\pi^2} (1-\alpha^2) k_{\rm B}T \sum_n 
    \int_0^{2\pi} d\theta \int_0^\infty pdp  \frac{1}{D_1^2 D_2^2},
\end{align}
with $D_1 = i\varepsilon_n+\mu +(\alpha\cos{}\theta -1)p$ 
and  $D_2 = i\varepsilon_n+\mu +(\alpha\cos\theta +1)p$. 
Using
\begin{align}
    \frac{1}{D_1 D_2} &= \frac{1}{2p} \left( \frac{1}{D_1} - \frac{1}{D_2} \right), \cr
    \frac{1}{p D_1} &= \frac{1}{D_0} \left( \frac{1}{p} - \frac{\alpha\cos\theta-1}{D_1} \right), \cr
    \frac{1}{p D_2} &= \frac{1}{D_0} \left( \frac{1}{p} - \frac{\alpha\cos\theta+1}{D_2} \right), 
\end{align}
with $D_0 = i\varepsilon_n+\mu$, we can show 
\begin{align}
    \frac{p}{D_1^2 D_2^2} 
    =&\frac{1}{4p} \left[ \frac{1}{D_1^2} + \frac{1}{D_2^2} -\frac{1}{p} 
    \left\{ \frac{1}{D_1} - \frac{1}{D_2} \right\} \right] \cr
    =&-\frac{1}{4D_0} \left( \frac{\alpha\cos\theta-1}{D_1^2} 
    + \frac{\alpha\cos\theta+1}{D_2^2} \right) \cr
    &-\frac{\alpha\cos\theta}{4D_0^2} \left( \frac{\alpha\cos\theta-1}{D_1} - \frac{\alpha\cos\theta+1}{D_2} \right).\notag \\
\end{align}
Using this relation, the $p$-integral in $\chi$ 
can be carried out, which leads to
\begin{align}
    \chi = \frac{e^2 v_{\rm F}^2}{24 \pi^2} (1-\alpha^2) k_{\rm B}T \sum_n 
    \frac{1}{D_0^2} \int_0^{2\pi} d\theta F(\theta, \alpha),  
\end{align}
with
\begin{align}
    &F(\theta,\alpha) = 2+\alpha\cos\theta \biggl[ 
    \frac{1}{2} \ln \frac{(\alpha\cos\theta-1)^2}{(\alpha\cos\theta+1)^2} \cr
    &-\frac{i\pi}{2} {\rm sign}(\alpha\cos\theta-1) +\frac{i\pi}{2} {\rm sign}(\alpha\cos\theta+1)
    \biggr].
\end{align}

It is easy to see that the imaginary part in $F(\theta,\alpha)$ does not contribute 
to $\chi$ even if $\alpha>1$. The integral of the term with 
logarithm should be carried out with care.
When $\alpha<1$, using the integration by parts, we obtain
\begin{widetext}
\begin{align}
    &\int_0^{2\pi} d\theta \frac{\alpha\cos\theta}{2} 
    \ln \frac{(\alpha\cos\theta-1)^2}{(\alpha\cos\theta+1)^2} \cr 
    &= \alpha\sin\theta \ln \frac{(\alpha\cos\theta-1)^2}{(\alpha\cos\theta+1)^2} \biggl|^\pi_0
    +\int_0^\pi d\theta \left( \frac{2\alpha^2\sin^2\theta}{\alpha\cos\theta-1} 
    -\frac{2\alpha^2\sin^2\theta}{\alpha\cos\theta+1} \right) \cr
    &=\int_0^\pi d\theta \left( -4 + \frac{2(\alpha^2-1)}{\alpha\cos\theta-1} 
    -\frac{2(\alpha^2-1)}{\alpha\cos\theta+1} \right) \cr
    &=-4\pi +4\pi \sqrt{1-\alpha^2}.
    \label{eq:PartialInt}
\end{align}
\end{widetext}
As a result, $\chi$ becomes
\begin{align}
    \chi = \frac{e^2 v_{\rm F}^2}{6\pi} (1-\alpha^2)^{3/2} k_{\rm B}T \sum_n 
    \frac{1}{D_0^2},  
\end{align}
for $\alpha<1$, which is equivalent to Eq.~(\ref{eq:Chi2DDirac3}). 

On the other hand, when $1<\alpha$, $\alpha\cos\theta -1$ or $\alpha\cos\theta+1$ 
in the logarithm vanishes when $\theta=\theta_0$ where $\theta_0$ satisfies 
$\sin\theta_0 = \sqrt{\alpha^2-1}/\alpha$.
To avoid the singularities in the integration by parts, we perform a different 
integration from (\ref{eq:PartialInt}) as
\begin{align}
    &\int_0^{2\pi} d\theta \frac{\alpha\cos\theta}{2} 
    \ln \frac{(\alpha\cos\theta-1)^2}{(\alpha\cos\theta+1)^2} \cr 
    &= \left (\alpha\sin\theta - \sqrt{\alpha^2-1} \right) \ln \frac{(\alpha\cos\theta-1)^2}{(\alpha\cos\theta+1)^2} \biggl|^\pi_0 \cr
    &+\int_0^\pi d\theta \left (\alpha\sin\theta - \sqrt{\alpha^2-1} \right) 
    \left( \frac{2\alpha\sin\theta}{\alpha\cos\theta-1} 
    -\frac{2\alpha\sin\theta}{\alpha\cos\theta+1} \right) \cr
    &=4\sqrt{\alpha^2-1}\ln \frac{\alpha-1}{\alpha+1} 
    - \int_0^\pi d\theta \frac{4\alpha \sin\theta}{\alpha\sin\theta+\sqrt{\alpha^2-1}} \cr
    &=-4\pi.
    \label{eq:PartialIntMod}
\end{align}
As a result, $\chi$ vanishes for $\alpha>1$.

\section{Equivalence of the two formulae of 
orbital magnetic susceptibility 
in Eqs.~(\ref{eq:fukuyama_formula}) and 
(\ref{eq:Piechon_formula}) \label{appB}}

First, we can show that
\begin{equation}
\frac{\partial}{\partial k_\mu} G_{\bm k}(\epsilon)
=G_{\bm k}(\epsilon) v_{\bm k}^\mu G_{\bm k}(\epsilon),
\end{equation}
with $\mu=x, y$. Using this relation and a trick, 
we see that
\begin{align} 
 \Theta(\epsilon)= 
 \frac{1}{V}\sum_{\bm{k}}  
&\mathrm{Tr} 
\biggl[\frac{2}{3} v^x_{\bm{k}} G_{\bm{k}}(\epsilon)v^y_{\bm{k}} G_{\bm{k}}(\epsilon)v^x_{\bm{k}} G_{\bm{k}}  (\epsilon)v^y_{\bm{k}} G_{\bm{k}}(\epsilon) \cr
&+ \frac{1}{3} v^x_{\bm{k}} \left( \frac{\partial}{\partial k_y} G_{\bm{k}}(\epsilon) \right) v^x_{\bm{k}} G_{\bm{k}}(\epsilon) v^y_{\bm{k}} G_{\bm{k}}(\epsilon)\biggr]. 
\end{align}
Then, we perform the integration by parts in the 
second term assuming that the boundary contributions 
of the Brillouin zone vanish. For lattice models, the vanishing of the surface terms can be proven~\cite{Mizoguchi2024}.
Furthermore, in the case where 
$v_{\bm k}^{xy} =0$ holds, 
the second term becomes
\begin{align} 
 \frac{1}{V}\sum_{\bm{k}}  
\mathrm{Tr} 
\biggl[
&-\frac{2}{3} v^x_{\bm{k}} G_{\bm{k}}(\epsilon)v^x_{\bm{k}} G_{\bm{k}}(\epsilon)v^y_{\bm{k}} G_{\bm{k}}  (\epsilon)v^y_{\bm{k}} G_{\bm{k}}(\epsilon) \cr
&-\frac{1}{3} v^{x}_{\bm{k}} G_{\bm{k}}(\epsilon) v^{x}_{\bm{k}} G_{\bm k}(\epsilon) v^{yy}_{\bm{k}} G_{\bm{k}} \biggr]. \label{eq:AppB3}
\end{align}
In the similar way, the second term in 
Eq.~(\ref{eq:AppB3}) can be transformed as
\begin{align} 
&\frac{1}{V}\sum_{\bm{k}}  
\mathrm{Tr} 
\biggl[
-\frac{1}{3} v^{x}_{\bm{k}} 
\left( \frac{\partial}{\partial k_x} G_{\bm{k}}(\epsilon) \right) v^{yy}_{\bm{k}} G_{\bm{k}} \biggr] \cr
=&\frac{1}{V}\sum_{\bm{k}}  
\mathrm{Tr} 
\biggl[
\frac{1}{3} v^{x}_{\bm{k}} 
 G_{\bm{k}}(\epsilon) v^{yy}_{\bm{k}} G_{\bm{k}}(\epsilon) v_{\bm k}^x G_{\bm k}(\epsilon) \cr
&+\frac{1}{3} v^{xx}_{\bm{k}} 
 G_{\bm{k}}(\epsilon) v^{yy}_{\bm{k}} G_{\bm{k}}(\epsilon) \biggr]. 
\label{eq:AppB4}
\end{align}
The first term on the right-hand side is the same 
as the left-hand side by using the cyclic nature 
of the trace. 
As a result, we can see that the second term in 
Eq.~(\ref{eq:AppB3}) is equal to
\begin{equation}
\frac{1}{V}\sum_{\bm{k}}  
\mathrm{Tr} 
\biggl[\frac{1}{6} v^{xx}_{\bm{k}} 
 G_{\bm{k}}(\epsilon) v^{yy}_{\bm{k}} G_{\bm{k}}(\epsilon) \biggr]. 
\end{equation}
Combining the above formulae, we can see that
Eq.~(\ref{eq:Piechon_formula}) gives the same 
result with Eq.~(\ref{eq:fukuyama_formula}) 
because $v_{\bm k}^{xy}=0$.

\bibliographystyle{apsrev4-2}
\bibliography{orb}

\end{document}